\documentclass[a4paper,12pt]{article}

\usepackage{graphicx}

\title{Propagation of thermonuclear flame in SNIa}

\author{S. I. Glazyrin and S. I. Blinnikov\\
        Institute for Theoretical and Experimental Physics,\\ Moscow, Russia}
\date{}
      
\begin{document}

\maketitle

\begin{abstract}
  The propagation of thermonuclear flame in presupernovae Ia
  is considered. Front parameters are obtained, some speculations on
  front stability are presented.
\end{abstract}                

\section{Introduction}

The problem of thermonuclear flame propagation in supernovae Ia still stands. Full
hydrodynamic simulation requires knowledge of small scale parameters of
flame: its normal propagation velocity, instability regimes.
In this paper we show instability manifestations using a toy
model. In literature there is no clear understanding whether flame
front is stable or pulsates under instability \cite{BychkovLiberman}.
In the second part we carry out full hydrodynamical simulations of
flame and obtain flame parameters for the range of densities, a
similar analysis was performed in \cite{TimmesWoosley}.

\section{Toy Model}

Let us consider a simple model for evolution of temperature $T$ and reagent
fraction $c$\footnote{The model was proposed by P.V. Sasorov
(ITEP)}:
\begin{equation}
  \partial_t T=\kappa\partial_x^2 T+\omega_0
  c\Theta(T-T_0),~~\partial_t c=-\omega_0 c\Theta(T-T_0),
\end{equation}
where $\Theta$ is a theta-function (a step-function). The system
models deflagration burning in solid propellants because two main
physical processes that drive slow front are presented in it:
thermoconductivity and burning itself.
Medium in supernovae is gaseous, but when flame propagates in the
centre of the white dwarf, in dense
matter ($\rho\sim 10^8\div 10^9$ g/cm$^3$) density jump is low, so
hydrodynamical effects are small and evolution matches burning of solid
medium. Moreover, at Lewis number ${\rm Le}\gg 1$, the process of
burning in supernovae in general is similar to that described to our system.
The choice of burning rate function is
explained below.

A stationary wave must obey boundary conditions:
\begin{equation}
  t=0,~x\rightarrow\infty:~~T=0,~~c=1,~~\partial_x
  T=0,~~~~t=0,~x\rightarrow -\infty:~~c=0.
\end{equation}
The system can be simplified by redefinition of $x$ to put $\kappa=1$.
We are searching for the wave front, so every quantity depends only on
$\xi=x-vt$. Due to translation invariance we put $\xi=0:~T=T_0$ (the
point of center of flame).
The system can be easily solved:
\begin{eqnarray}
  \xi>0:&~~c=1,~~T=T_0e^{-v\xi},\nonumber\\
  \xi<0:&~~c=e^{\omega_0\xi/v},~~T=1-\frac{\omega_0}{(\omega_0/v)^2+\omega_0}e^{\omega_0\xi/v},\\
  &v=\sqrt{\frac{1-T_0}{T_0}\omega_0}\nonumber.
\end{eqnarray}
For more simplification we put $v=1$, that means $\omega_0=T_0/(1-T_0)$.
Let us finally write down the simplified system and its solution:
\begin{equation}
  \partial_t T=\partial_x^2 T+\omega_0
  c\Theta(T-T_0),~~\partial_t c=-\omega_0 c\Theta(T-T_0),
  \label{sys:simpl}
\end{equation}
\begin{eqnarray}
  \xi>0:&~~c=1,~~T=T_0e^{-\xi},\nonumber\\
  \xi<0:&~~c=e^{\omega_0\xi},~~T=1-\frac{1}{\omega_0+1}e^{\omega_0\xi}.
  \label{sys:simpl_solution}
\end{eqnarray}

The stability of such a system under small perturbations can be easily
considered analytically:
\begin{equation}
  T=T_{\rm n.p.}+T_{\rm p},~~c=c_{\rm n.p.}+c_{\rm p},
\end{equation}
\begin{equation}
  T_{\rm p}=e^{pt}f(\xi),~~c_{\rm p}=e^{pt}g(\xi).
  \label{eq:pert_pt}
\end{equation}
After some calculations \cite{GlazyrinSasorov} the following result
could be obtained: the system is stable when $\omega_0<6$, and
perturbations grow exponentially when $\omega_0>6$.

\begin{figure}[!htb]
\begin{center}
\includegraphics[height=10cm,angle=270]{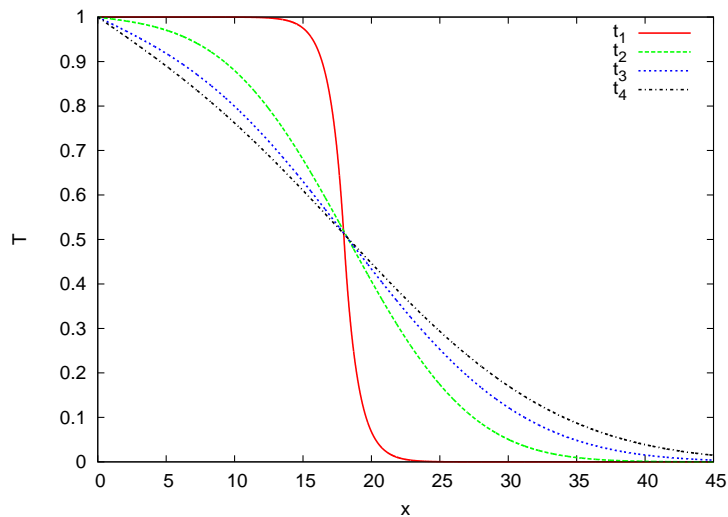}
\caption{\label{fig:Tt_80_1}Front positions for $\omega_0=8$ at
  different time moments: $t_0<t_1<t_2<t_3$.}
\end{center}
\end{figure}

Such a system can be easily numerically simulated and full evolution
of unstable regime could be obtained. The task is set as follows:
\begin{equation}
  c|_{t=0}=c_{\rm theor},~~T|_{t=0}=T_{\rm theor},
\end{equation}
where $c_{\rm theor}$ and $T_{\rm theor}$ are defined in
(\ref{sys:simpl_solution}) with center at $x_{c}$ and bound conditions:
\begin{equation}
  c|_{x=0}=0,~~c|_{x=L}=1,~~T|_{x=0}=1,~~T|_{x=L}=0.
\end{equation}
So we set the exact analytical solutions as initial conditions and watch
their evolution. The Table \ref{tab:toy} shows results of simulations. 
Solutions could be split into two groups: ``flame'' and
``therm''. ``flame'' -- is the evolution as stationary flame front with
constant velocity, this regime exists when $\omega_0<6$ (according
to the table of results). ``therm'' describes flame decay, like evolution under
thermoconductivity without burning (example of such evolution is shown
in Fig. \ref{fig:Tt_80_1}). Zero burning rate when
$T<T_0$ forbids flame ignition after ``therm'' regime, so it clearly
cuts only unstable evolution.

\begin{table}
\begin{center}
\begin{tabular}{|c|c|c|c|}
  \hline
  $\omega_0$ & $v$ & comm. \\
  \hline
  1.0 & 1.000 & flame \\
  4.0 & 0.996 & flame \\
  5.5 & 1.006 & flame \\
  5.8 & 1.010 & flame \\
  6.0 & 1.019 & flame \\
  6.1 & -- & therm \\
  7.0 & -- & therm \\
  8.0 & -- & therm \\
  9.0 & -- & therm \\
  \hline
\end{tabular}
\end{center}
\caption{\label{tab:toy}Numerical simulation. $v$ -- measured front velocity.}
\end{table}

The analytical predictions are in a very good agreement with numerical
simulation of the model. So this model and its
modifications may be used for theoretical study of unstable flame
fronts.

\section{One-dimensional flame properties}

Let us consider full hydrodynamical evolutions of flame in
presupernova Ia. Our goal is to study flame acceleration and
deflagration to detonation transition \cite{ImshennikKhokhlov1991ab}. Typical parameters of medium in the center of WD
are the following \cite{HillebrandtNiemeyer}: $T\sim 10^9$ K, $\rho\sim 10^9$ g/cm$^3$, chemical
composition -- $^{12}$C$+^{16}$O. Let us
suppose that only $^{12}$C remains in chemical composition. For given
conditions the following relations hold: Pr $\ll 1$, Le $\gg
1$. It means that thermoconductivity is the only diffusion mechanism
that matters in this case. The coefficient of thermoconductivity is
the sum of two parts: electron conductivity \cite{YakovlevUrpin} and radiative
conductivity \cite{Iben}. We consider the only one
nuclear reaction for approximation (also this approach gives ability to
study physical effects):
\begin{equation}
  ^{12}{\rm C}+^{12}{\rm C}\rightarrow ^{24}{\rm Mg}^*.
\end{equation}
Here Mg$^*$ means the excitation state, it decays through 3 channels: with n,
with p, with $\alpha$; and we take into account the sum of rates, which
could be found in \cite{CaughlanFowler}.
Its caloricity is $q_1=5.5\cdot 10^{17}$ erg/g.
The reaction is the first in the network, it occurs between two
highly charged nuclei (it is one of the slowest), so we could suppose
that it determines the whole rection rate.
Futher burning could be indroduced in our model by changing
caloricity. For burning up to $^{56}$Ni it will be $q_2=9.2\cdot
10^{17}$ erg/g.
Nuclear screening should be taken into account because $\Gamma =
\langle E_{\rm coul}\rangle/kT\sim 1$.

\begin{figure}[!htb]
\begin{center}
\includegraphics[height=11cm,angle=270]{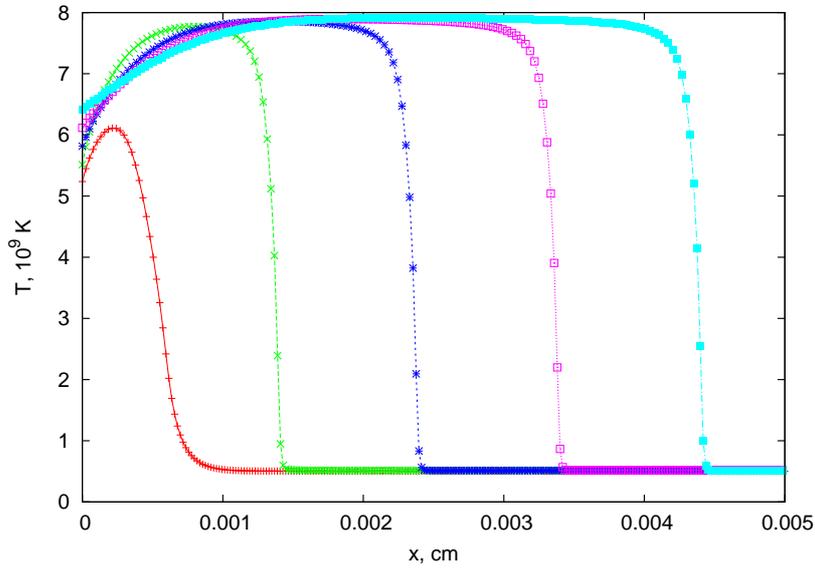}
\caption{\label{fig:T_x_t} Sequental temperature distributions (for
  different physical time).}.
\end{center}
\end{figure}

The problem is formulated in the following way: a full one-dimensional
hydrodynamical system of equations with thermoconductivity and nuclear
reactions is considered. It is solved by our numerical code FRONT
(where an implicit numerical scheme with a Newton iterations solver is used
\cite{SamPopov}).
Initial region of calculation is filled uniformly with $^{12}$C at given $T_0$ and
$\rho_0$. Right wall should be free for stream. Left wall is heated by
the linear law $T=T_0+t(T_1-T_0)/\tau$. Where $\tau$ obeys $\tau\gg
L/c_{s}$ ($L$ -- the size of region of interest, $c_s$ -- sound
speed). Such conditions lead to deflagration wave ignition by the hot
wall.
The sequental flame positions are shown on Fig. \ref{fig:T_x_t}.
Table \ref{tab:real} shows results of normal flame speed determination
for different initial density $\rho_0$.

\begin{table}
\begin{center}
  \begin{tabular}{|c|c|c|c|c|c|c|}
    \hline
    $\rho_0$, g/cm$^3$ & Cal. & $T_{\rm max}$, $10^9$ K & $\rho_u$, g/cm$^3$ & $\rho_b$, g/cm$^3$ & $v_n$, km/s & $\Delta x_{\rm fr}$, cm \\
    \hline
    $2\cdot 10^8$ & Mg & 6.8 & $2.05\cdot 10^8$ & $1.35\cdot 10^8$ & 222 & $5\cdot 10^{-4}$ \\
    & Ni & 7.9 & $2.12\cdot 10^8$ & $1.16\cdot 10^8$ & 460 & $4\cdot 10^{-4}$ \\
    \hline
    $7\cdot 10^8$ & Mg & 9.0 & $7.27\cdot 10^8$ & $5.38\cdot 10^8$ & 888 & $4.6\cdot 10^{-5}$ \\
    & Ni & 10.8 & $8.08\cdot 10^8$ & $5.07\cdot 10^8$ & 1950 & $5.3\cdot 10^{-5}$ \\
    \hline
    $2\cdot 10^9$ & Mg & 11 & $2.10\cdot 10^9$ & $1.67\cdot 10^9$ & 1880 & $1.1\cdot 10^{-5}$ \\
    & Ni & 13 & $2.37\cdot 10^9$ & $1.62\cdot 10^9$ & 3450 & $2.8\cdot 10^{-5}$ \\
    \hline
  \end{tabular}
\end{center}
\caption{\label{tab:real}Measured deflagration flame front parameters
  for reaction $^{12}$C+$^{12}$C.}
\end{table}

It should be emphasized here that all velocities obtained are correct only for
one reaction in network $^{12}$C+$^{12}$C. Use of full nuclear network
changes the speed radically \cite{BlinnikovGlazyrin}.

\section{Conslusions}

The toy model was presented for several puproses: first it clearly shows
how instability of flame front manifests: it
leads to front destruction, second, the model could be used in
theoretical speculations and in academic studies.
Full hydrodynamic simulations shows stable
front propagations at all considered densities (with no evolution
shown by ``therm'' regime in toy model), so we could state that
thermonuclear flame front is stable in all density range from $2\cdot
10^8$ g/cm$^3$ to $2\cdot 10^9$ g/cm$^3$, the reason for that should
be explored. Flame front paratemeters
were obtained in hydrodynamical simulations. Front velocities differs
greatly from determined in \cite{TimmesWoosley}. The reason is in very
simple nuclear network: only one reaction is taken into account (for
detailes see \cite{BlinnikovGlazyrin}). The work is supported partly by
Federal Program ``Scientific and pedagogical specialists of innovation
Russia'' contract number 02.740.11.0250, partly by the Russian
Foundation for Basic Research grant RFBR 10-02-00249-a, by 
SNSF SCOPES project No.~IZ73Z0-128180/1, and ``Dynasty'' foundation.

\end{document}